# Vanadium dioxide circuits emulate neurological disorders

**Authors:** J. Lin[1,2], S. Guha[1,2], and S. Ramanathan*[3,4]

**Affiliations:**

[1]Center for Nanoscale Materials, Argonne National Laboratory, Lemont, IL 60439 USA

[2]Institute for Molecular Engineering, University of Chicago, Chicago, IL 60615 USA

[3]School of Materials Engineering, Purdue University, West Lafayette, IN 47907 USA

[4]School of Electrical and Computer Engineering, Purdue University, West Lafayette, IN 47907, USA

*Correspondence to:

Prof. S. Ramanathan

Addr.: 701 W. Stadium Ave, Armstrong Hall of Engineering, Purdue University, IN 47907

Phone: 765-496-0546

**Article type:**

Original Research

**Abstract:**

Information in the central nervous system (CNS) is conducted via electrical signals known as action potentials and is encoded in time. Several neurological disorders including depression, Attention Deficit Hyperactivity Disorder (ADHD), originate in faulty brain signaling frequencies. Here, we present a Hodgkin-Huxley model analog for a strongly correlated $VO_2$ artificial neuron system that undergoes an electrically-driven insulator-metal transition. We demonstrate that tuning of the insulating phase resistance in $VO_2$ threshold switch circuits can enable direct mimicry of neuronal origins of disorders in the central nervous system. The results introduce use of circuits based on quantum materials as complementary to model animal studies for neuroscience, especially when precise measurements of local electrical properties or competing parallel paths for conduction in complex neural circuits can be a challenge to identify onset of breakdown or diagnose early symptoms of disease.

# I. Introduction

Action potentials (AP) are generated in neurons and propagated to other neurons via synapses (Hodgkin and Huxley, 1952; Kandel, 2012). The frequency of the spikes carries information and is critical for brain function. How frequently neurons spike for a given stimulus and whether or not they are able to travel without losing signal strength dictate normal vs. abnormal brain function (Bartzokis, 2005; Salinas and Sejnowski, 2001; Wulff et al., 2009). Alteration in neural oscillations caused by abnormal excitation of action potential has been found to play an important role in a number of neurological disorders. Many research works on molecular neurophysiology have suggested the correlation of pathologically altered action potential excitability. Various neurological diseases are briefly summarized as follows.

Wu. et. al. reported that the dysfunctional calcium channel in mutant mouse model is associated with the hypokalemic periodic paralysis which is a form of paroxysmal weakness that occurs in motor neuron disease (Wu, 2012). Research has shown that the Alzheimer's disease can occur due to disruption of neuronal excitability. As examples, Chakroborty et. al. showed that increase of frequency and amplitude of AP due to certain protein channel dysregulation results in excitability impairment, with which the Triple Tg expression model is developed (Chakroborty et al., 2009; Santos et al., 2010). Drug addiction is also strongly related to abnormal excitation of action potential. Kourrich et. al. revealed the relationship between drug addiction and brain activity. They found that neurons subjected to certain dose of Cocaine will fire about 30% faster at high input current, and 200% faster at low input current (Kourrich et al., 2015). *Global Burden of Disease Study* revealed that major depression was the second largest cause of disability (estimated by the loss of productivity from the disease) and it affected approximately 300 million people worldwide in 2010 (Vos et al., 2012). Friedman et. al. showed that the midbrain dopamine neurons have played important in certain depressions. When the dopamine neuron (in mice) fire rate increases by 50% (from 1.6 Hz to 2.4 Hz), the social interaction (measured by a special experiment, see reference) dropped by about 60% (Friedman et al., 2014). Bipolar disorder, also known as manic depression, has been studied in monkeys and it is found that the mental disorder is related to the prefrontal cortical neurons firing and signaling at the molecular level (Birnbaum et al., 2004). Another example in the context of understanding and curing neurological syndromes is the research on neuropathic pain. Researchers sought for treatment for pain by treatments to tune the neuron oscillation frequency (Campbell and Meyer, 2006). These studies and references contained in them strongly indicate the crucial role of controlling the spiking frequencies and action potential generation in order-disorder transitions in biological neural circuits. Besides these examples, other neurological disorders that are resulted from pathologically-altered brain signaling frequencies also include neuromuscular diseases (Hutchison et al., 2004; Nelson and Valakh, 2015; Younger, 1999), ADHD (Brennan and Arnsten, 2008) and etc..

Understanding their origins and the mechanisms to minimize damage to neural pathways is a principal area of study in neuroscience. However, diagnosis of neurological disorder at the molecular level is challenging (Brown et al., 2004). One widely adopted method for neurophysiological measurements is the multiple-electrode recording of the electrical signal of AP spikes in brain tissue (Brown et al., 2004). To-date, neural recording experiments usually involve invasive probing (Kinney et al., 2015) and the *in vivo* measurements are mostly carried out on small animals such as mice (Barry, 2015; Schulz et al., 2014). Artificial circuits that mimic desired signal propagation characteristics along neurons and can provide parametric information on normal-abnormal signaling transitions from electrical properties of circuit components could be valuable in evaluating or directing animal studies. Here, we propose understanding electrical behavior of neurons and neurological disorders via synthetic circuits comprised of a strongly correlated oxide $VO_2$ that undergoes an electrically-driven *insulator-metal transition* (*IMT*).

Oxides have been studied for electronic devices such as resonant tunneling diodes, single-electron transistors, and steep slope switches (Mannhart and Schlom, 2010; Vitale et al., 2015). Among these emerging oxide-based electronic device concepts, phase changing artificial neurons has primarily focused on applications in neuromorphic computing to mimic the leaky-integrate-fire function (Lin et al., 2016; Pickett et al., 2013; Tuma et al., 2016). Here, we present a *Hodgkin-Huxley (HH) model analog* for the intrinsic properties of a solid-state material, $VO_2$. The strongly correlated $VO_2$ artificial neuron system can undergo an electrically driven insulator-metal transition akin to the excitable membrane in the biological neuron. Changes in composition of the material synergistically modifies the ground state resistivity, IMT strength defined as resistance ratio in the two phases as well as the threshold voltage required for initiating a phase change. Such material property is designed to capture the Intrinsic Membrane Excitability (IME) in biological neurons, which refers to a neuron's propensity for generating action potential at a given input. Building on this fundamental concept, we demonstrate neuronal function mimicking a vast range of neuron types found in animal brains and simulate an archetypal monosynaptic circuit (e.g. the *knee-jerk reaction*). Long term, our results may help in creating artificial systems to generate knowledge about thresholds for onset for brain disorders due to neuronal malfunction.

## II. Materials and Methods

$VO_2$ thin films of 200 nm thickness were deposited on $SiO_2$/Si by reactive sputtering at 775 K. The stoichiometry and IMT transition strength in $VO_2$ is controlled by the oxygen partial pressure in the sputtering chamber. The insulator-metal transition occurs at a critical voltage $T_c$. IMT transition strength ($R_{ins}/R_{met}$) is defined by the ratio of high resistance state ($R_{ins}$, measured at room temperature) and low resistance state ($R_{met}$, measured at above critical transition temperature). In $VO_2$, $T_c$ is 67°C. The low resistance state is taken at 120°C that is significantly higher than $T_c$. Our film growth experiments have shown controllable thermal IMT strength variation from $R_{ins}/R_{met}$ >$10^5$ to $R_{ins}/R_{met}$=1 (complete loss of IMT characteristic) (Ha et al., 2013; Lin et al., 2016; Zhou and Ramanathan, 2015). $R_{ins}$ and $R_{met}$ are respectively the resistivity for the insulating state and metallic state, and is characterized by temperature-dependent Hall measurement.

Three $VO_2$ samples were used for device fabrication and artificial neuron circuit testing. Their IMT strength is respectively 20, 750, and $2 \times 10^5$. For device fabrication, electron-beam lithography (EBL) is first used to define the length of the $VO_2$ device, L, as shown in Fig. 1A. L is in the channel direction and it varies from 50 nm to 17 μm. Ti/Au of thickness 5/100 nm were evaporated to form electrical contacts to the $VO_2$. A "neck-down" design for the contact were used as illustrated in Fig. 1B. The neck-down device drives only a small volume of the $VO_2$ into transition and reduces the voltage required to trigger the transition (Lin et al., 2017).

All experiments were carried out at room temperature. The DC sweeping and current-clamp are both performed using Keysight B1500A. Waveforms for the current-clamp experiment were acquired by Keysight Digital Oscilloscope DSO9104A. For DC sweeping, a current compliance is set to 5 mA. For current-clamp response, the IMT device is protected by a series resistance through a circuit board so that the measurement can be repeated reliably by avoiding excess heating and burnout.

Fig. 1A shows the schematic of the $VO_2$ devices under DC probing and Fig. 1B is the top view of the lateral $VO_2$ device with a "neck-down" layout. The "neck-down" layout is used to minimize the volume of $VO_2$ that undergoes transition (Lin et al., 2017). It leads to lower critical transition voltage ($V_c$) and lower power in the switching operation. The smallest spacing between two contacts is L= 200 nm for the device being studied in this work. Fig. 1C shows a measured current versus voltage characteristic under DC condition. A hysteresis sweep is performed and the switch between insulator-state and metal-state is reversible if the operation satisfies the safe criteria introduced in (Lin et al., 2017).

Excessive bias stress to the $VO_2$ device can result in non-reversible damage to the material which is manifested in a permeant change in critical transition voltage under DC measurement. This can happen when the device is subjected to a bias outside the safe operating criteria. Two forms of non-reversible damages are shown in Fig. 2. Two forward DC sweeps (in the positive direction) are carried out consecutively in one device. Fig. 2A shows an increase in $V_c$ caused by an increase of the HRS resistance (+$\Delta R$). The current drops over the whole range of applied voltage in the second sweep. Fig. 2B shows a reduction in $V_c$ which is the indication of a drop in the HRS resistance (-$\Delta R$). The current is higher over the span of applied voltage.

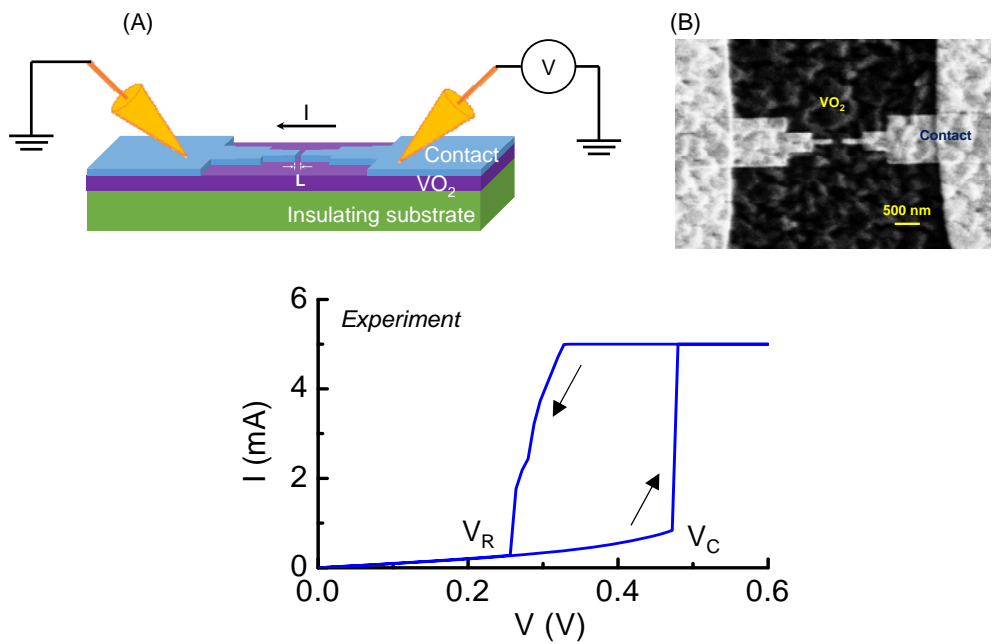

**Fig. 1. Experimental details for the $VO_2$ device.** (A) The schematic of the $VO_2$ devices under DC probing. (B) The top view of the lateral $VO_2$ device with a "neck-down" layout. The "neck-down" layout leads to lower forward critical transition voltage ($V_c$) and lower power in the switching operation. (C) The typical current versus voltage characteristic in DC measurement. The voltage sweep is in the sequence of forward (0 to 0.6 V) and reverse (0.6 to 0 V) direction. The switch is reversible. The forward critical voltage and reverse voltage are denoted.

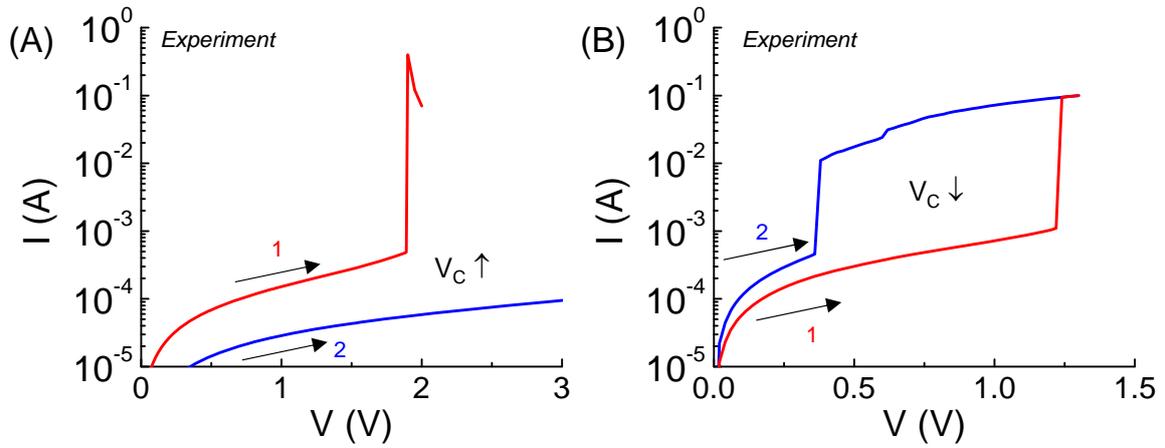

**Fig. 2.** VO$_2$ device under excessive DC stress experiencing non-reversible change in critical transition voltage. Two consecutive sweeps are applied to the VO$_2$ device with the first sweep stresses the device. (A) Increase in $V_c$ is caused by an increase of HRS resistance, $+\Delta R$. (B) Reduction in $V_c$ is the indication of a drop in HRS resistance, $-\Delta R$.

# III. Results

Fig. 3 shows schematic of a biological neuron and an analogous $VO_2$ neuron. The membrane of the biological neuron (Fig. 3A) comprises of an insulating phospholipid bilayer that separates the intracellular and extracellular fluids, and protein channels that control the permeation of various ions. As described in the Hodgkin-Huxley model, the neuron membrane is equivalent to a parallel combination of membrane capacitance, $C_m$, and transmembrane conductance, $G_m$. $G_m$ is the sum of various ion channels conductance and it can go through a reversible *insulator-to-metal transition* depending on the voltage across the membrane. An input stimulus can trigger a train of action potentials (AP) that is a temporary reversal of the polarity across the neuron membrane. The AP propagates along the axon through which information is transmitted. In the central nervous system (CNS) such as in the brain and spinal cord, the neuron is myelinated – the myelin sheath surrounds the axon of the neuron cells and promotes rapid signal transmission. Experimentally, the neuron can be stimulated with an input current $I_{in}$. The output AP waveforms depend on the input current, the electrical and geometric parameters of the cell, and the environment such as temperature. The HH model and its parameters are described later in Sec. II.A.

The $VO_2$ device and analogous $VO_2$ neuron circuit are shown in Fig. 3B. The $VO_2$ is well-known for its reversible *insulator-metal transition* proximal to room temperature. Joule heating in two-terminal devices can locally drive the phase change rapidly. This property can be exploited to demonstrate highly nonlinear switches (Lin et al., 2017; Son et al., 2011) and artificial neurons (Lin et al., 2016; Pickett et al., 2013; Tuma et al., 2016). The $VO_2$ artificial neuron is a circuit that comprises, a minimum of, only two components, the capacitor $C_o$ and the conductor (i.e. resistor) $G_{VO}$ as shown in Fig. 3B. When the input stimulus $I_{in}$ starts, the $VO_2$ neuron exhibits an oscillatory behavior similar to that in the biological neuron. The model and experiment for standalone $VO_2$ devices are discussed respectively in Sec. II.B and Sec. II.C. The $VO_2$ neuron circuit is described in Sec. II.D.

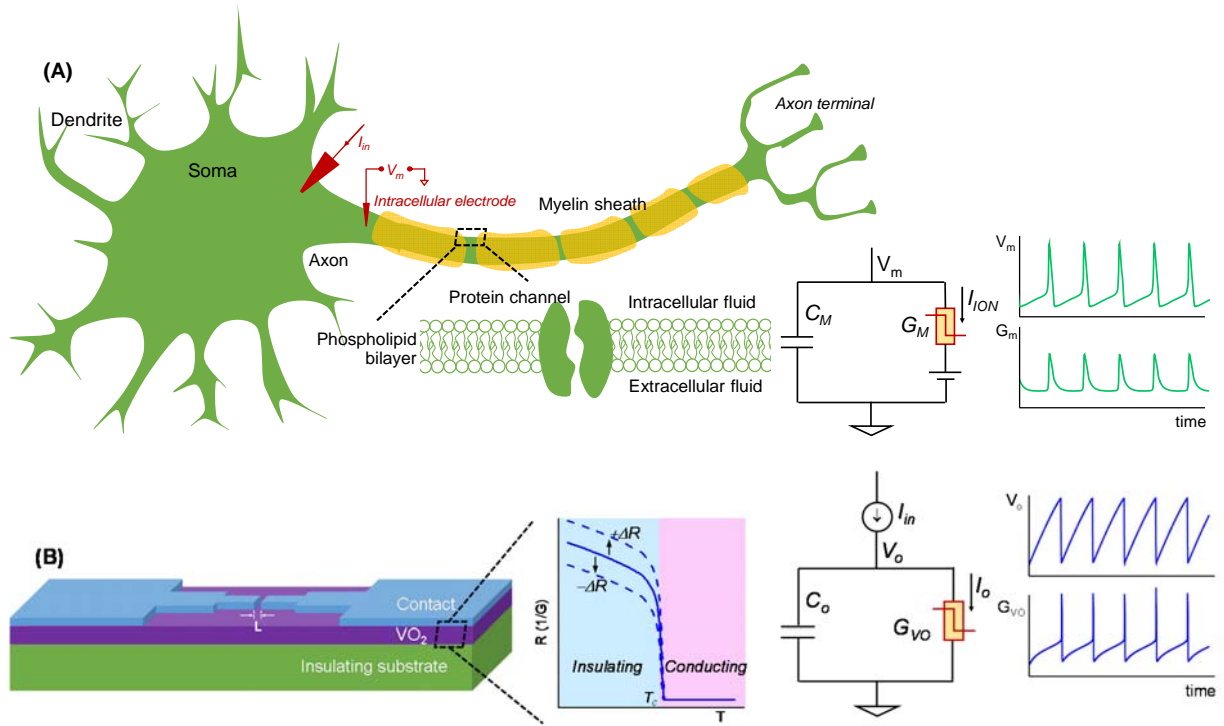

**Fig. 3. The biological neuron and analogous VO$_2$ neuron.** (**A**) The membrane of the biological neuron can be viewed as a parallel connection of a membrane capacitance (C$_M$) and a membrane conductance (G$_M$) that can go through *insulator-to-metal transition* under stimulus. The membrane is polarized at the resting potential due to the different ionic concentration in the intracellular and extracellular fluids. When the neuron is subjected to a steady current clamping, a continuous action potential (AP) is generated, in which the trans-membrane potential (V$_M$) and the membrane conductance (G$_M$) oscillate. The AP can propagate along the axon and transmit signal to the other connected neurons. The myelin sheath surrounding the axon of some neuron cells can enhance the speed at which impulses propagate. (**B**) A lateral VO$_2$ device and a capacitor are used to construct the VO$_2$ artificial neuron circuit. The VO$_2$ material exhibits a reversible electrothermal *insulator-to-metal transition*. This state change is used to mimic the biological neuron. At constant current input, the VO$_2$ neuron output node and VO$_2$ conductance oscillate, similar to that of the biological neuron. The insulating-state resistance can be changed when VO$_2$ degrades, and this feature is utilized to model spike-timing related neural disorders. Here +ΔR represents an increase of resistance and −ΔR represents a drop in resistance.

# A. Hodgkin-Huxley (HH) Model for Biological Neuron

The complete circuit schematic for a patch of the neuron membrane with the HH model is illustrated in Fig. 4. The conduction occurs via three channels: The $Na^+$ channel, the $K^+$ channel and the leakage channel.

The basic mechanisms in the HH model contains ion transport and transmission lines for Action Potential (AP) propagation. The key equations are shown in Eqs. 1-3. Eq. 1 relates the membrane current density $I_{IN}$ to membrane potential $V_m$. The area-normalized membrane capacitance is $C_m$. Two ion channels with the leaky conductance are included in the model. Their conductance is denoted as $G_{Na}$, $G_K$ and $G_L$. The Nernst equilibrium potential in Eq. 2 relates extracellular and intracellular ion concentrations, respectively denoted as $C^i$ and $C^o$. Through Eq. 2, the Nernst equilibrium potentials $V_{Na}$ and $V_K$ can be obtained for the given $Na^+$ and $K^+$ concentrations. The molar gas constant R and Faraday's constant F are physical constants. Finally, the propagation of AP along z direction is described by the core conductor equation in Eq. 3. It couples the voltage and current along a cylindrical cell where the resistances per unit length inside and outside the cell are respectively $r_i$ and $r_o$. The cylindrical cell has diameter a. The baseline values of the parameters in the HH model are listed in Table 1. The HH neuron model is constructed in Matlab.

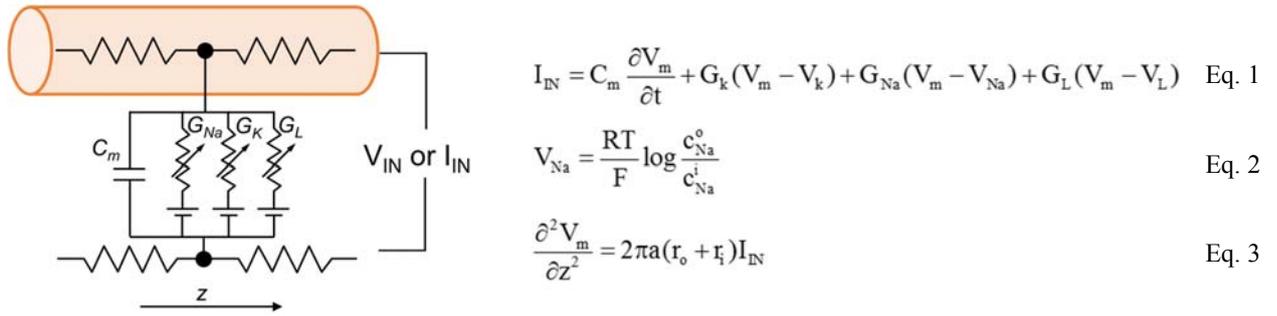

$$I_{IN} = C_m \frac{\partial V_m}{\partial t} + G_k(V_m - V_k) + G_{Na}(V_m - V_{Na}) + G_L(V_m - V_L) \quad \text{Eq. 1}$$

$$V_{Na} = \frac{RT}{F} \log \frac{c^o_{Na}}{c^i_{Na}} \quad \text{Eq. 2}$$

$$\frac{\partial^2 V_m}{\partial z^2} = 2\pi a (r_o + r_i) I_{IN} \quad \text{Eq. 3}$$

**Fig. 4. Full schematic of biological neuron that contains two ion channels, leaky capacitive membranes.** The equations the form the Hodgkin-Huxley model are shown in the Eq. 1-3.

Table 1: Baseline input parameters for the Hodgkin-Huxley model

| Parameter | Symbol | Value |
|---|---|---|
| Conductance of Sodium channel | $G_{Na}$ (mS/cm$^2$) | 120 |
| Conductance of Potassium channel | $G_K$ (mS/cm$^2$) | 36 |
| Leakage conductance | $G_L$ (mS/cm$^2$) | 0.3 |
| Extracellular Sodium concentration | $c^o_{Na}$ (mmol/l) | 500 |
| Intracellular Sodium concentration | $c^i_{Na}$ (mmol/l) | 50 |
| Extracellular Potassium concentration | $c^o_K$ (mmol/l) | 20 |
| Intracellular Potassium concentration | $c^i_K$ (mmol/l) | 400 |
| Membrane capacitance | $C_m$ (µF/cm$^2$) | 1 |

Fig. 5 shows the results from the HH model with a current clamp. The input is two discrete current pulses. The current-clamped neuron is subjected to the deposition of charge from each of the pulse. The deposited charge of one pulse is merely enough to trigger one neuron firing. The solid red lines are the results for pulse width of 3 ms, and the dashed black lines are for pulse width of 6 ms. The resultant AP profiles are identical for different pulse width, even the long pulse (dashed black) has deposited twice as many charges as the short pulse (solid red). The additional charges are not integrated because it falls into the "refractory period". A new integration cycle starts only after the neuron resets itself.

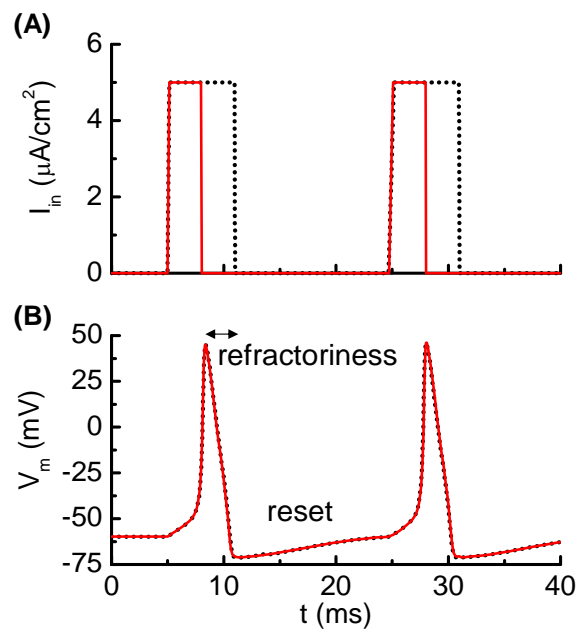

**Fig. 5. Simulated results from the Hodgkin-Huxley model.** (A) The neuron is clamped to pulse current input, and (B) cross membrane potential. The red solid lines and black dot lines are two different inputs that result in the same firing patterns. The extra current inputs are not being integrated to the membrane capacitance due to the existence of post-refractory period.

## B. Model for VO₂ Device

The model for electrothermal insulator-metal transition is discussed in this section. The basic form of heat equation is a parabolic partial differential equation (Eq. 4) that describes the relationship of temperature variation in a given volume over time. Eq. 4 assumes an isotropic and homogeneous medium in a 3-dimensional space and zero heat flux. The 3D heat transport is simplified by the quasi-1D assumption that the temperature variation perpendicular to the current transport direction along y and z is much smaller than that in the transport direction along x. This is illustrated in Fig. 6 with the current flowing in x direction, and the heat equation is converted to Eq. 5. Under this assumption, the temperature for a given segment at location x and time t is obtained as $T(x, t)$. The parameters describing the IMT property are given as follows, specific heat capacity $C_{th}$, density $\rho_o$, and thermal conductivity K.

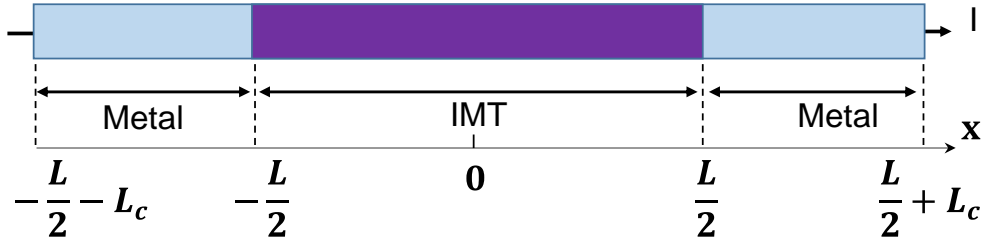

**Fig. 6. Full schematic of an IMT device with length L between two metal contacts.** The device is discretized into segment of dx, etch with its own temperature and resistivity. Heat conduction is along x direction while convective heat loss through the side wall.

Table 2. Input parameters for the coupled electrical-thermal model IMT model.

| Parameter | Symbol | Value |
|---|---|---|
| Thermal conductivity | K (W/K-m) | 6 |
| Specific heat capacity | C* (J/K-kg) | 690 |
| Effective convective heat transfer coefficient | h (W/K-m²) | 10 |
| High resistivity state | $\rho_{rH}$ (Ω-m) | $10^{-3}$ |
| Low resistivity state | $\rho_{rL}$ (Ω-m) | $10^{-5}$ |
| Density | $\rho_d$ (kg/m³) | 4x10³ |
| Cross sec' area | A (m²) | 1x10⁻¹² |
| Length | L (m) | 5x10⁻⁷ |
| Ambient temperature | $T_a$ (K) | 300 |
| Series resistance | $R_S$ (Ω) | 70 |

$$C_{th}\rho_o \frac{\partial T}{\partial t} = K(\frac{\partial^2 T}{\partial x^2} + \frac{\partial^2 T}{\partial y^2} + \frac{\partial^2 T}{\partial z^2})$$
(Eq. 4)

$$C_{th}\rho_o \frac{\partial T}{\partial t} = K\frac{\partial^2 T}{\partial x^2}$$
(Eq. 5)

There are two origins of heat flux. Firstly, Joule heating results in incoming heat flux to the medium. The power generated by Joule heat follows Ohm's law, and for a unit volume it is:

$$P_1 = \frac{I^2 \rho_r}{A^2}$$
(Eq. 6)

where I is the total current through the IMT, A is the cross sectional area, and $\rho_r$ is the temperature-dependent resistivity of the IMT. Secondly, the outgoing heat flux is generated by convective heat loss, modeled by the effective convective heat transfer coefficient h. Here h is assumed to be a constant and is independent of the IMT temperature. Besides h, the heat loss through convection for a unit volume is also related to the ambient temperature $T_a$, and IMT's surface to volume ratio $L_p/A$ where $L_p$ is the cross sectional perimeter. The power dissipated through side wall heat convection is:

$$P_2 = \frac{h \cdot L_P (T - T_a)}{A}$$
(Eq. 7)

Taking into account the heat fluxes, the differential equation for heat transfer is shown in Eq. 8.

$$C_{th}\rho_o \frac{\partial T}{\partial t} = K\frac{\partial^2 T}{\partial x^2} + \frac{I^2 \rho_r}{A^2} - \frac{h \cdot L_P (T - T_a)}{A}$$
(Eq. 8)

As shown in Fig. 6, the IMT with length L is connected to two metal contacts with length $L_c$. As the boundary condition, $L_c$ is assumed to be long enough so that the value of $L_c$ has negligible impact – $L_c$ should be significantly longer than the heat diffusion length. Eq. 8 is solved using a numerical method: forward difference for the time domain and central difference for spatial domain. In the spatial domain, the IMT bar is discretized into segment of length dx. Each segment has its resistivity $\rho_r(x)$. The total IMT resistance $R_{IMT}$ is obtained by integrating the resistance of all segments:

$$R_{IMT} = \int_{-L/2}^{L/2} \frac{\rho_r(x)}{A} dx$$
(Eq. 9)

The current through the IMT can then be obtained by

$$I = \frac{V}{R_{IMT} + R_S}$$
(Eq. 10)

Where $R_s$ is the series resistance. The IMT resistivity as a function of temperature follows a look-up table of resistivity versus temperature as measured in the experimental VO$_2$ device. A typical example of the resistivity versus temperature is shown in Fig. 3B, which is characterized by the insulator-state resistivity $\rho_H$, the metal-state resistivity $\rho_L$, and the critical transition temperature ($T_c$). The baseline values of other physical parameters are listed in Table 2.

### C. VO$_2$ Neuron Circuit

The complete VO$_2$ neuron circuit is shown in Fig. 7. This is one of the simplest artificial neuron circuits that has been reported, which comprises of only two or three elements. Despite its simplicity, it exhibits striking similarity to the biological neuron (Fig. 3A). To investigate the many unexplored functions of the artificial neuron, particularly its connection to neurological diseases, a physical neuron model is imperative.

The model is focuses on the material properties of VO$_2$ that emulate biological neuron functions. A series resistance $R_s$ is added in series with the VO$_2$ for two reasons. First, it limits the current when the VO$_2$ device transitions to the metallic state, and ensures reliable switching. The safe operating design follows the theoretical guideline derived in (Lin et al., 2017). Second, it converts the output current to an output voltage which is measured by the oscilloscope. Each spiking event includes the following four steps: integration, fire, refractoriness and reset as discussed in Fig. 8.

One example of the simulation result is shown in Fig. 9. Two discrete current pulses are fed to the input node of the neuron circuits. The deposited charge from each pulse is enough to fire the neuron once. The pulse duration is 0.9 μs for the red solid lines and 1.8 μs for the black lines (Fig. 9A). After each neuron firing, the VO$_2$ stays in low resistance state for a finite period (Fig. 9B). During this period, the input current are directly drained to ground through the VO$_2$. The charge is not integrated. As a result, inputs with two pulse durations generates the same firing patterns (Fig. 9C). This results show similar post-firing refractoriness as the biological neuron (Fig. 5).

Post-firing refractoriness is another important feature in both biological and VO$_2$ neurons. A second AP is difficult to be produced immediately following the occurrence of an AP when the cell is regarded to be refractory (Weiss, 1996). Following each firing, the VO$_2$ element remains at a temperature above the critical temperature for a time, $\sim\tau_{th}+\tau_{el}$, where $\tau_{th}$ and $\tau_{el}$ are the thermal and electrical time constants respectively. $\tau_{th}$ is related to the thermal mass and heat dissipation. For the electrical time constant $\tau_{el}$, it is given by $R_{met}C_o+R_sC_o$. $R_{met}$ is the metallic-state resistance and $R_s$ is the series resistance. Usually, $R_{met}$ is much greater than $R_s$ in normal operation (Lin et al., 2017). Therefore, the refractory period is the same as the pulse width ($\tau_w$) and is given by $R_sC_o$. During this period the VO$_2$ element remains in metallic state and new input charge is continuously discharged without being integrated in the capacitor $C_o$. The quantity $\tau_r$ therefore defines the restitution time. Our coupled electrothermal model has captured this process and can be used to design the restitution time in the VO$_2$ neuron circuit. Subsequently, the VO$_2$ element resets and starts another integrate-and-fire cycle. The steps exactly mimic the electrically excitable membrane in neuron cells.

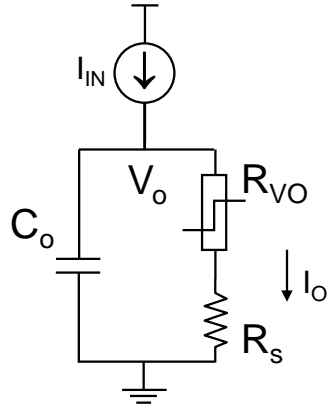

**Fig. 7. The complete VO$_2$ neuron circuit.** The whole circuit contains three elements: a capacitor as well as the VO$_2$ device with a sensing resistor in series. The output current is sensed by the sensing resistor. The voltage at the capacitor node is denoted as V$_o$. It is also the input node for the injected current.

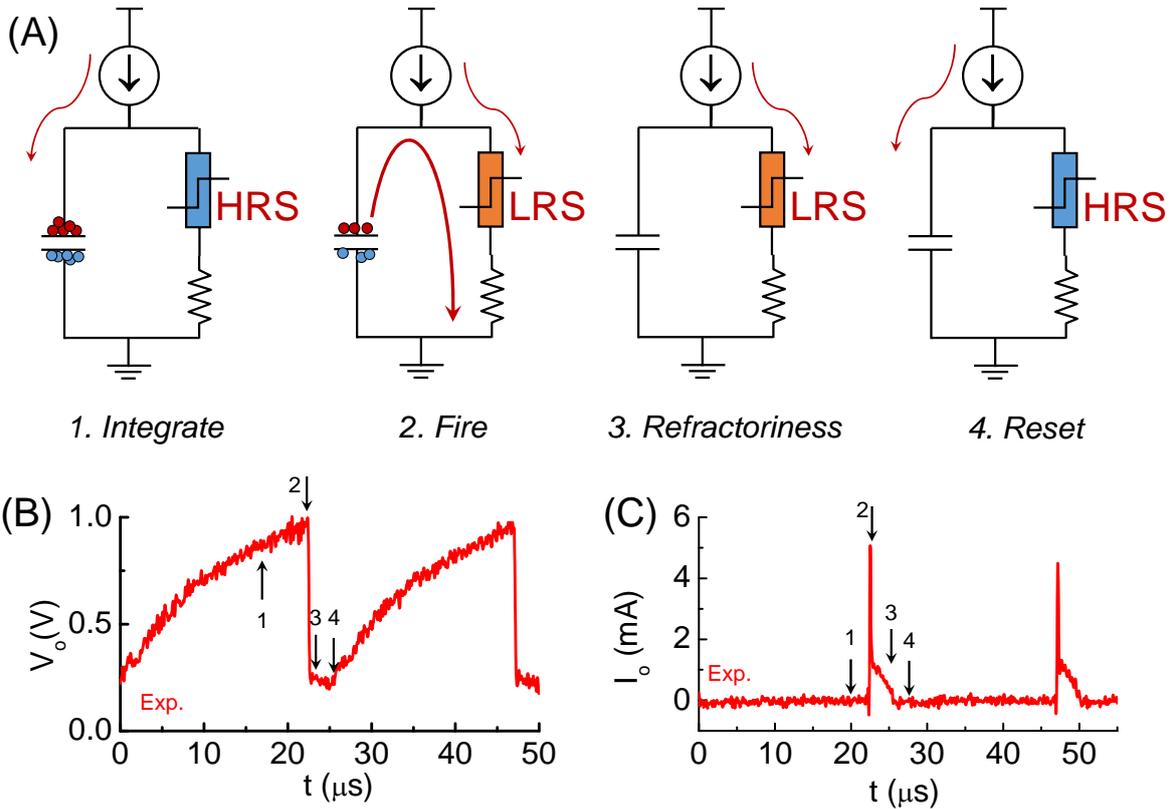

**Fig. 8. The four stages in one spike cycle in the VO$_2$ neuron and the corresponding experimental output waveforms.** At the integration stage, current is integrated to the capacitance. Very small current goes through the resistor at the right branch as the VO$_2$ is at HRS. Voltage at across the capacitor V$_o$ is increasing. When V$_o$ reaches V$_c$, VO$_2$ become metallic and it discharges the capacitor. An instantaneous large current spike appears at the output. V$_o$ drops sharply. This is stage 2, fire, which is followed by stage 3, refractoriness (refractory period). In stage 3 the VO$_2$ remains in its LRS for some time. Any input current will be drained to ground without integrating to the capacitor. After the refractory period, the neuron reset and is ready for another spike cycle.

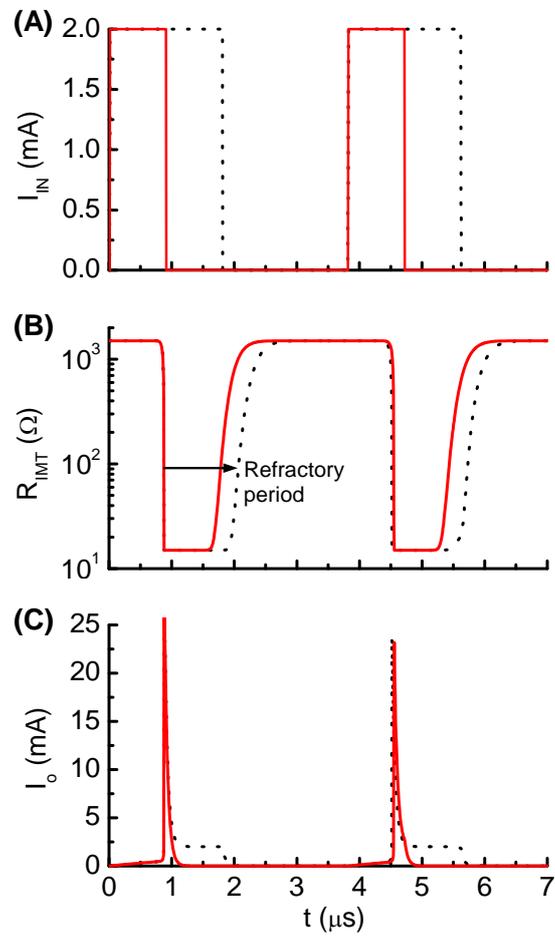

**Fig. 9. Simulated waveform for current-clamp VO$_2$ neuron**. (A) Input current, (B) VO$_2$ resistance and (C) Output current. The red solid lines and black dot lines are two different inputs that result in the same firing patterns. The extra current inputs are not being integrated to the capacitor due to the existence of post refractory period. This property directly mimics the biological neuron.

# IV. Discussion

### A. Healthy vs Degenerative Neurons, and Their $VO_2$ Analogy

In biological neurons, the inter-spiking interval (ISI) is defined as the time interval between two adjacent spikes (Fadool et al., 2011; Okubo et al., 2015). The spiking frequency is the reciprocal of ISI. The AP recorded as a function of time is shown for a healthy neuron (Fig. 10A) along with two abnormal neurons (Figs. 10 B and C), simulated with the HH model. A pathological change of the action potential firing frequency can lead to neurological and psychological disorders. For instance, a decrease in AP frequency is tied to CNS depression and cognitive dysfunction (Friedman et al., 2014). In contrary, an increase in firing frequency is responsible for seizures, pain, ADHD, and anxiety (Wulff et al., 2009). Therapeutic treatment can be designed to restore AP frequency according to the dysfunction mechanisms.

Similar characteristics can be observed in the $VO_2$ neurons. The simulation results for three $VO_2$ neurons are shown in Fig. 10. Fig. 10D is the $VO_2$ neuron for baseline reference, and Figs. 10 E and F are the cases where the HRS resistance is modified. ISI for the $VO_2$ neuron is defined in the same way as for the case of a biological neuron. The AP frequency reduces if the $VO_2$ undergoes a $+\Delta R$ degradation, and vice versa. Analytically, the value for ISI can be derived from the $VO_2$ neuron parameters as $t_{ISI}=C_o V_c/I_{in}$ where $V_c$ is the critical voltage to trigger an insulator-to-metal transition in the $VO_2$ device under DC I-V measurement and $I_{in}$ is the input current. $V_c$ is related to the HRS resistance of the $VO_2$ device. Definition of $V_c$ is illustrated in Fig. 2. Experimentally, we have observed the change of $V_c$ ($\Delta V_c$) due to electrical-stress-induced resistance degradation in the $VO_2$ (Fig. 7). $\Delta V_c$ be positive or negative depending on the degradation mechanism. Positive $\Delta V_c$ indicates an increase in the $VO_2$ HRS resistance, and vice versa.

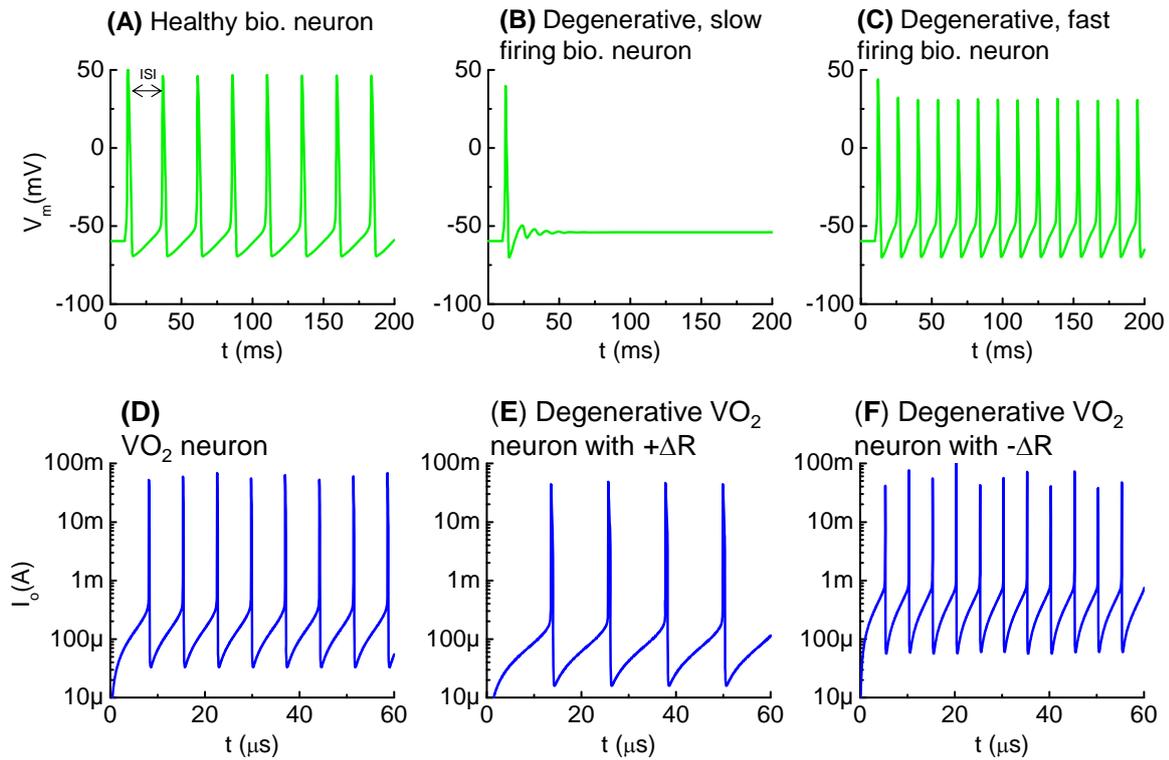

**Fig. 10. Healthy vs degenerative fast-firing and slow-firing neurons, and VO₂ analogy.** (**A**) A healthy biological neuron generates AP under a constant current stimulus. The time between two adjacent spikes is termed as the inter-spike interval (ISI). (**B**) A degenerated neuron stimulated by the same input generates AP at shorter ISI. (**C**) A degenerated neuron generates AP at longer ISI. (**D**) Simulation using the VO₂ neuron model, the intact VO₂ neuron exhibits oscillatory behavior at a constant input current. (**E**) Simulation of a case where the VO₂ device is degraded and its insulating state resistance increases (+ΔR). Such degenerative VO₂ neuron results in longer ISI. (**F**) Simulation of a case where the VO₂ device is degraded by decreasing its insulating state resistance increases (-ΔR). The leakier VO₂ neuron results in shorter ISI. (**A-C**) are simulated by the HH model. (**D-F**) are simulated from the VO₂ neuron model.

The pathologically-altered spike timing is linked to other serious degenerative diseases. For example, certain neuromuscular disorders and motor neuron disease (MND) are resulted from the ionic leakage of degenerating membrane and increase of rest conductance (Priori et al., 2002; Younger, 1999). The electrical breakdown of the myelin sheath is one origin for a leaky membrane. Excessive leakage in the membrane makes weak muscles (Wu, 2012). The HH model for biological neuron shows that a neuron fails to fire when conductance is significantly increased (Figs. 11 A and B). AP generation in a healthy neuron is accompanied by an insulator-to-metal transition in the membrane. However, no transition can be observed in a leaky membrane at the same current input since the charge cannot be integrated effectively.

The increase of conductance in a degenerative, leaky neuron can be modeled in a straightforward manner in the VO₂ circuit. The resistance versus temperature of the VO₂ device is normalized to the low resistance state, $R_{met}$. Figs. 11 C and D show two neurons at a fixed $I_{in}$: (a), VO₂ neuron with $R_{ins}/R_{met}=10^5$ fires regularly and demonstrates an insulator-to-metal transition (b) while the neuron with $R_{ins}/R_{met}=10^3$ fails to fire. The change of $VO_2$ properties is shown in Fig. 11E.

To provide a systematical perspective on the design of $VO_2$ neurons to mimic the corresponding neural disorder, Fig. 11F collectively illustrates the impact of HRS resistance and input current on neuron functions. The value of inter-spike interval depends on $R_{ins}$ and $I_{in}$ and is shown as a contour plot. The dark red color is the case where the combination of low input current and small $R_{ins}$ results in failure in spike generation ($t_{ISI} \to \infty$). At a given $I_{in}$, $t_{ISI}$ decreases as $R_{ins}$ drops. When $R_{ins}$ drops to a critical value, the $VO_2$ neuron fails to fire (Fig. 11G).

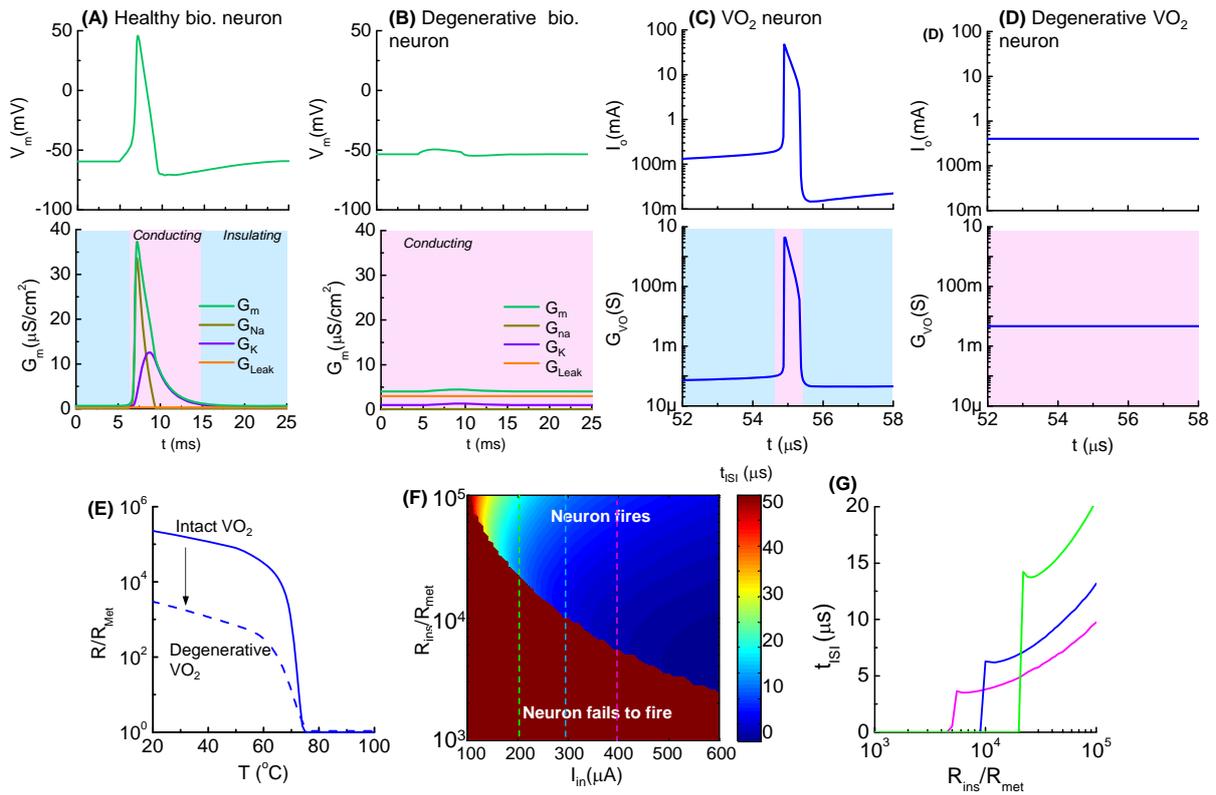

**Fig. 11. Healthy vs degenerative leaky neurons, and $VO_2$ artificial neurons.** (**A**). Healthy biological neuron shows an insulator-to-metal transition during one spike event. $G_{Na}$, $G_K$, and $G_{Leak}$ are respectively the $Na^+$ conductance, $K^+$ conductance and leakage conductance through the membrane, while $G_M$ is the the sum of the conductance. (**B**) Degenerative biological neuron with excess leakage $G_{Leak}$, while $G_K$ and $G_{Na}$ remain unchanged. No AP spike is observed. (**C**) In one spike of the $VO_2$ neuron, the $VO_2$ device goes through an insulator-to-metal transition. (**D**) Degenerative $VO_2$ neuron with excess leakage (-ΔR). (**E**) The resistance as a function of temperature normalized to the metallic state resistance ($R_{ins}/R_{met}$) illustrates the reduction of resistance of the insulating state by ~100. (**F**) Contour of ISI shows its dependency on material properties ($R_{ins}/R_{met}$) and input stimulus ($I_{in}$). Reduction of insulating-state resistance narrows the neuron operating region for a given input stimulus. (**G**) Three cut lines across $I_{in}$=200, 300 and 400 μA in the contour plot (F). (A-B) are simulated by the HH model. (C-G) are the simulated results from the $VO_2$ neuron model.

The AP pulse width is another distinctive characteristic related to timing in different kinds of mammalian central neurons (Bean, 2007). A short pulse and a long pulse in biological neurons are respectively illustrated in Figs. 12 A and B, with the $t_w$ change by 10X. The pulse width can be simulated in the VO2 neuron by altering the LRS resistance, $R_{met}$, in the VO2 devices according to $t_w=C_oR_{met}$. The VO2 neuron with short pulses of 0.2 μs and 2 μs are shown in Figs. 12 C and D. Fig. 12E compares the biological neurons and VO2 neurons (both simulations and experiments). The AP pulse width can range from a few 100 μs to 10 ms which can be matched by the VO2 neuron with appropriate capacitance.

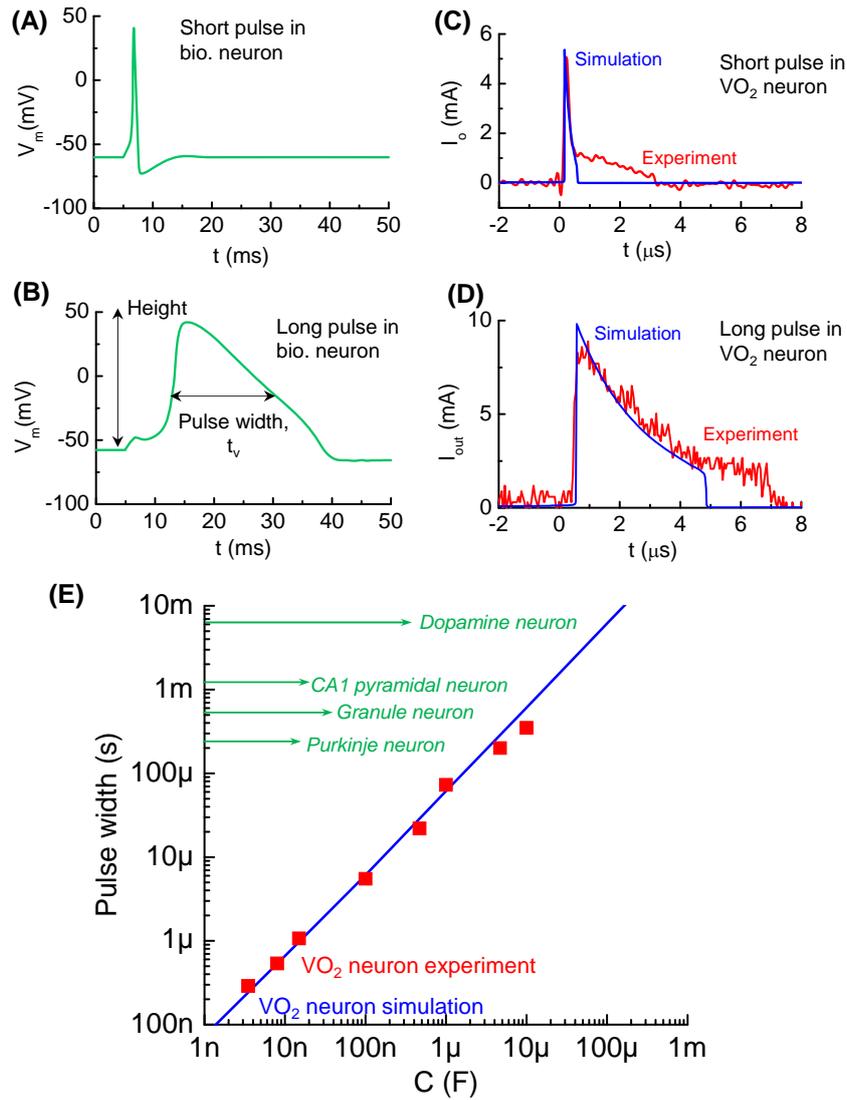

**Fig. 12. Diversity in AP pulse width across biological neurons, and VO2 artificial neuron analogy.** (**A**). AP with short pulse width in a biological neuron. The value of $t_w$ is taken at full width half maximum. (**B**) AP with long pulse width. (**C**) VO2 neuron with short pulse of 0.2 μs. Experiment and simulation show good agreement. The pulse width control is achieved by changing resistance in the circuit (**D**) VO2 neuron with long pulse of 2 μs. (**E**) Pulse width of the VO2 neuron and the range spans biological neuron studies reported in the neuroscience literature.

## B. Monosynaptic Neuron Circuit

We further extend this concept to two-stage cascading neuron circuits in Fig. 13. The circuit is a modeling system for the monosynaptic motor neuron in muscular tissue that is responsible for certain motion responses such as the *knee-jerk reaction* that is a model system in neuroscience. Neuron 2 (in red) is the receptive neuron that is driven by Neuron 1 (in blue). Neuron 2 can either fire or not fire depending on the output waveform of Neuron 1 as well as the synaptic resistor $R_x$.

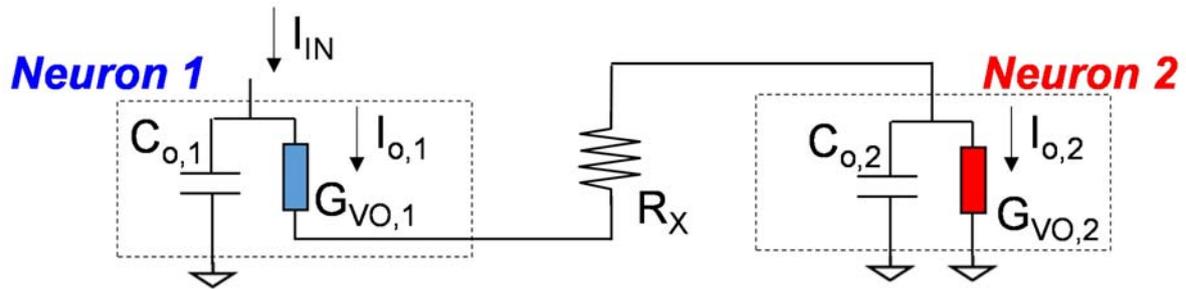

**Fig. 13. VO$_2$ monosynaptic neuron circuit.**

We emulate a monosynaptic circuit that corresponds to the well-known *knee-jerk reaction* used to monitor responses in nerves (Kandel, 2012) as shown in Fig. 14. In the two cases for Figs. 14 A and B, Neuron 2 is the same while the HRS resistance of VO$_2$ in Neuron 1 are different. Both neurons are initially at rest. Their temperatures are at equilibrium with the environment and is below the critical transition temperature $T_c$. At t=0, a current is injected to Neuron 1 (see Fig. 9). The neuron 1 in case A is intact with high $R_{ins}/R_{met}$, The output current as a function of time in Fig. 14A shows the spike events for Neurons 1 and 2. The separation of spikes indicate the reaction time ($t_{diff}$=0.6 μs) for the signal to propagate between the two VO$_2$ neurons. The neuron 1 in case B has a lower HRS resistance. Premature spike in Neuron 1 results in a weak spike and it cannot trigger a spike in Neuron 2 (Fig. 14B).

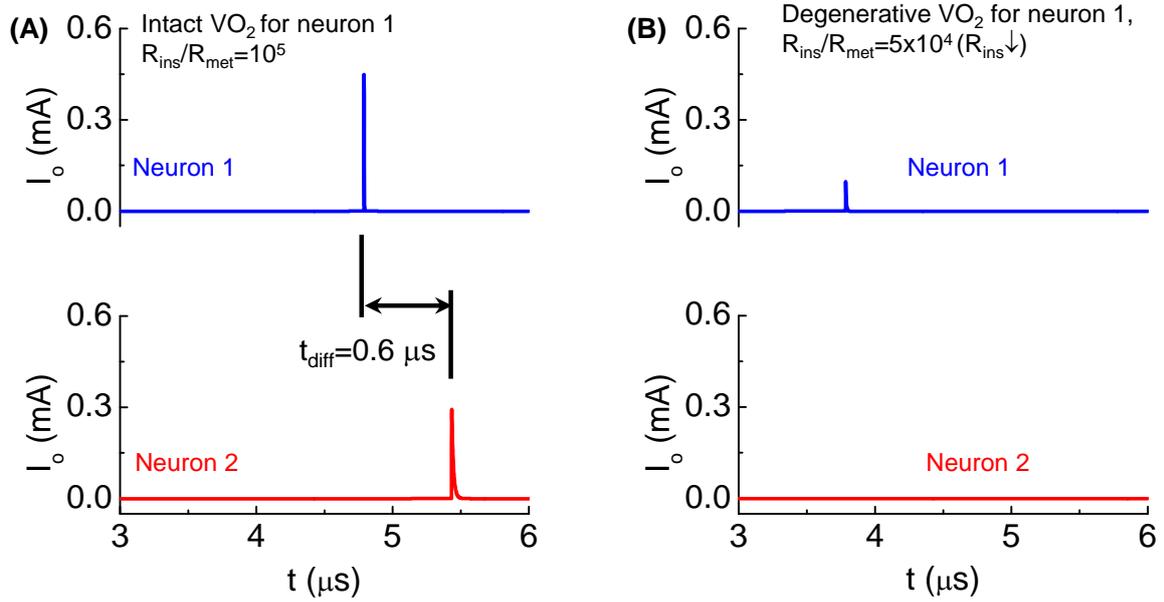

**Fig. 14. Demonstration of degenerative Neuron 1 leading to the failed signal reception for Neuron 2 in a monosynaptic circuit.** (**A**) Output current as a function of time shows the spike events for the intact case. (**B**) Premature spike in Neuron 1 when its HRS resistance is reduced and Neuron 2 fails to spike.

## V. CONCLUSION

VO$_2$ based circuits can emulate neuronal function and disorders. By carefully varying the electrical properties of the ground state resistance of the artificial neuron, we can precisely identify thresholds for firing and signal propagation that present an analogy to neuronal activity in the brain. While the present study has focused on VO$_2$ as a model system, a vast range of threshold switching Mott semiconductors can further be explored in the future.

**Acknowledgments:**

This work was performed, in part, at the Center for Nanoscale Materials, a U.S. Department of Energy Office of Science User Facility under Contract No. DE-AC02-06CH11357. Aspects of the device work was supported by the National Science Foundation under grant 1640081, and the Nanoelectronics Research Corporation (NERC), a wholly owned subsidiary of the Semiconductor Research Corporation (SRC), through Extremely Energy Efficient Collective Electronics (EXCEL), an SRC-NRI Nanoelectronics Research Initiative under Research Task ID 2698.001. S.R. acknowledges the support by ARO W911NF-16-1-0289 and ONR N00014-16-1-2398. The authors acknowledge K. V. L. V. Achari for providing vanadium dioxide film samples.